\documentclass[twoside]{article}
\usepackage[a4paper]{geometry}
\usepackage[latin1]{inputenc} 
\usepackage[T1]{fontenc}
\usepackage{RR}

\RRdate{April 2013}

\usepackage{amsmath,amssymb,amsthm}
\usepackage{xspace}
\usepackage{graphicx}
\usepackage{color}
\usepackage{cite}

\usepackage{algorithm}
\usepackage[noend]{algorithmic}
\usepackage{longtable}

\usepackage{multirow}

\usepackage{array}
\usepackage{mdwmath}
\usepackage{mdwtab}
\usepackage[caption=false,font=footnotesize]{subfig}
\usepackage{stfloats}
\usepackage{eqparbox}

\usepackage{url}

\newcommand{\drmsim}{{DRMSim}\xspace}

\title{Feasibility study on distributed simulations of BGP}

\RRauthor{
  David Coudert\thanks{COATI Project, INRIA/I3S(CNRS/UNSA), France.
    {\tt david.coudert@inria.fr}}
  \and
    Luc Hogie\thanks{COATI Project, INRIA/I3S(CNRS/UNSA), France.
    {\tt luc.hogie@inria.fr}}
  \and
  Aur\'elien Lancin\thanks{COATI Project, INRIA/I3S(CNRS/UNSA), France.
    {\tt aurelien.lancin@inria.fr}}
      \and
  Dimitri Papadimitriou\thanks{Alcatel-Lucent Bell, Belgium.
    {\tt dimitri.papadimitriou@alcatel-lucent.com}}
      \and
  St\'{e}phane Perennes\thanks{COATI Project, INRIA/I3S(CNRS/UNSA), France.
    {\tt stephane.perennes@inria.fr}}
          \and
  Issam Tahiri\thanks{COATI Project, INRIA/I3S(CNRS/UNSA), France.
    {\tt issam.tahiri@inria.fr}}
}

\authorhead{D. Coudert \& L. Hogie \& A. Lancin \& D. Papadimitriou \& S. Perennes \& I. Tahiri }

\RRtitle{Etude de faisabilité de la distribution des simulations de BGP}

\RRetitle{Feasibility study on distributed simulations of BGP}

\titlehead{Feasibility study on distributed simulations of BGP}

\RRabstract{The Autonomous System (AS)-level topology of the Internet  that currently comprises 40k ASs, is growing at a rate of about 10\% per year. In these conditions, Border Gateway Protocol (BGP), the inter-domain routing protocol of the Internet starts to show its limits, among others in terms of the number of routing table entries it can dynamically process and control. To overcome this challenging situation, the design but also the evaluation of alternative dynamic routing models and their comparison with BGP shall be performed by means of simulation. For this purpose, DRMSim, a Dynamic Routing Model Simulator, was developed that provides the means for large-scale simulations of various routing models including BGP. By means of this discrete-event simulator, execution of path-vector routing, e.g. BGP, and other compact routing models have been successfully performed on network topologies comprising more than ten thousand (abstract) nodes. However, to simulate dynamic routing schemes like BGP, DRMSim needs enhancements to support current Internet size (40k ASs) and even more by considering its evolution (up to 100k ASs). This paper proposes a feasibility study of the extension of DRMSim so as to support the Distributed Parallel Discrete Event paradigm. We first detail the possible distribution models and their associated communication overhead. Then, we analyze the communication overhead of such a distributed simulator by executing BGP on a partitioned topology according to different scenarios. Finally, we conclude on the feasibility of such a simulator by computing the expected additional time required by a distributed simulation of BGP compared to its sequential simulation.}

\RRresume{La topologie d'Internet est constituée d'environ 40K systèmes autonomes (AS) et s'accroit de 10\% par année. Dans ces conditions, BGP (Border Gateway Protocol), le protocole de routage inter-domaine utilisé dans Internet commence à atteindre ses limites, plus particulièrement en terme de nombre d'entrées dans les tables de routage qu'il peut traiter dynamiquement. Afin de surmonter ce problème, la conception et l'évaluation de modèles de routage alternatifs et leur comparaison avec BGP doivent s'appuyer sur la simulation. Dans ce but, DRMSim, un simulateur de modèles de routage dynamiques, a été développé afin de fournir les moyens nécessaires pour la simulation à grande échelle de divers modèles de routage, BGP inclus. Grâce à ce simulateur à événements discrets, BGP qui est basé sur le routage par vecteurs de chemins et d'autres modèles de routage ont été simulés avec succès sur des topologies composées de plus de 10 000 noeuds. Cependant, pour simuler des schémas de routage dynamiques comme BGP, DRMSim doit être amélioré afin de fonctionner sur la topologie actuelle d'Internet composée d'environ 40K ASs, voir plus  si l'ont considère son évolution (jusqu'à 100K ASs). Dans ce papier nous proposons une étude de faisabilité pour étendre les fonctionnalités de DRMSim afin d'utiliser le concept de simulation parallèle et distribuée à événements discrets. Dans un premier temps, nous détaillons les divers modèles de distributions ainsi que le coût supplémentaire en terme de communication. Nous analysons ensuite ces coûts de communication pour un tel simulateur distribué lors de l'exécution de BGP sur une partition de la topologie suivant différents scénarios. Enfin, nous concluons notre étude sur la faisabilité d'un tel simulateur en relevant le temps supplémentaire nécessaire à une simulation séquentielle de BGP par rapport à celle distribuée.}

 \RRmotcle{Simulation à évènements discrets, simulation distribuée, réseaux, BGP}

 \RRkeyword{Discrete event Simulation,Distributed Simulation,Network,BGP}

\RRprojet{COATI}

\RCSophia

\begin{document}
\RRNo{}
\makeRR

\section{Introduction}

In \textit{distributed routing}, essential function of the Internet, each node independently executes a routing function that computes for any reachable destination name a loop-free routing path so that incoming messages directed to a given destination can reach it. As already reported twenty years ago (see ~\cite{rfc1287}), the evolution of the Internet routing system its underlying protocol, the Border Gateway Protocol (BGP) \cite{rfc4271} to its limits in terms of scalability and convergence properties. These limits result from the path-vector routing algorithm underlying BGP: i) memory consumption resulting from its stretch-1 routing paths; ii) its convergence properties resulting from path exploration events. Solving these issues requires to address multiple challenges altogether: i) the increasing size of the routing tables (measured in terms of memory space consumption) required to sustain the growing number of routing table entries resulting from the increasing number of routers and autonomous systems. 
; ii) the increasing communication cost or complexity (measured as the rate of exchange between nodes of routing messages in order for the routing function to properly operate) as the dynamics of the routing information exchanges between routers increasingly impacts the routing system convergence properties; iii) the increasing architectural complexity (measure of the complexity of a given architecture proportionally to its number of components and the number of interactions among components) of the Border Gateway Protocol (BGP) routing protocol itself, underlying the Internet routing system. Hence, routing research has focused on the investigation of new routing paradigms to address the above mentioned challenges altogether. 

On the other hand, investigation of new routing paradigms leads to a fundamental question. How to determine the performance and other properties of interest such as convergence of these new routing schemes on large scale dynamic topologies such as the Internet in order to determine suitable alternatives. Indeed, the simulation of stateful distributed routing protocols operating asynchronously over networks (of the order of tens of thousands of nodes) becomes a real issue at large-scale \cite{rob5}. To the best of knowledge of the authors, no simulator provides the means to measure and to characterize the performance and behavior of such routing schemes when applied to large networks ($> 10k$ nodes) and to compare them to BGP using the simulation environment. For this purpose, we propose in this paper to extend the capabilities of \textit{\drmsim} \cite{drmsim}, a discrete-event simulator which allows measuring the performance of routing models on large-scale networks. More precisely, we study the extension of this simulator in order to support the distribution of the routing model by partitioning the topology with respect to its properties and by extending the communication model in order to enable the distributed execution of the routing model. 

This paper is organized as follows. After describing in Section \ref{stateoftheart} the state-of-the-art in the domain of routing model simulation, we introduce \drmsim in Section \ref{simulator}. Then, we detail the distributed parallel discrete-event simulation paradigm and two distributed models together with their associated communication overhead. In Section \ref{execution}, we describe our simulation scenarios and execution environment followed by the simulation results as well as the network partition impact on the communication and their analysis. Finally, we conclude in Section \ref{conclusion}.
\section{State of the art}
\label{stateoftheart}

We have to distinguish three classes of simulators when it comes to routing: (routing) protocol simulators, routing configuration simulators, and (routing) model simulators.

Simulators dedicated to the performance measurement and analysis of the routing protocol (procedures and format) at the microscopic level. These can be further subdivided between dedicated BGP simulators and general simulators (which offer too execution of many other routing protocols). The ns ~\cite{ns} discrete-event simulator that relies on the BGP daemon from Zebra ~\cite{Zebra} belongs to the first sub-category. This daemon can be used to build realistic inter-domain routing scenarios but not on large-scale networks due to the low level execution of the protocol procedures. On the other hand, SSFNet ~\cite{SSFNET} discrete event simulator, relies on the implementation of the BGP protocol that was tailored and validated for the needs of a BGP-specific simulators. In SSFNet, a simulated router running BGP maintains its own forwarding table. It is thus possible to perform simulation with both TCP/IP traffic and routing protocols to evaluate the impact of a change in routing on the performance of TCP as seen by the end systems (hosts, terminals, etc.).

Simulators dedicated to simulation of BGP protocol specifics including the computation of the outcome of the BGP route selection process by taking into account the routers' configuration, the externally received BGP routing information and the network topology but without any time dynamics. These simulators can be used by researchers and ISP network operators to evaluate the impact of modified decision processes, additional BGP route attributes, as well as logical and topological changes on the routing tables computed on individual routers assuming that each event can be entirely characterized. Topological changes usually comprise pre-determined links and routers failures whereas logical changes include changes in the configuration of the routers such as input/output routing policies or IGP link weights. These simulators are thus specialized and optimized (in terms, e.g., of data structures and procedures) to execute BGP on large topologies with sizes of the same order of magnitude than the Internet since these simulators are not designed to support real-time execution. These simulators usually support complete BGP decision process, import and export filters, route-reflectors, processing of AS\_path attributes and even custom route selection rules for traffic engineering purposes, and BGP policies. Simulators like SimBGP ~\cite{SimBGP} or C-BGP~\cite{Quoitin05} belong to this category. These simulators are gradually updated to incorporate new BGP features but are complex to extend out of the context of BGP.

Simulators dedicated to the simulation of routing models, category to which \drmsim ~\cite{drmsim} belongs. Designed for the investigation of the performance of dynamic routing models on large-scale networks, these simulators allow execution of different routing models and enable comparison of their resulting performance. Simulators in this category consider models instead of protocols, meaning they do not execute the low level procedures of the protocol that process exact protocol formats but their abstraction. Thus these simulators require specification of an abstract procedural model, data model, and state model sufficiently simple to be effective on large-scale networks but still representative of the actual protocol execution. However, incorporating (and maintaining up to date) routing state information is technically challenging because of the amount of memory required to store such data. In practice, processing of individual routing states impedes the execution of large-scale simulations. \drmsim addresses this issue by means of efficient graph-based data structures. Moreover, by using advanced data structures to represent routing tables, \drmsim can still run simulation whose number of nodes exceeds ten thousands.

All simulators previously cited here above share many properties in common. Like \drmsim, they all rely on discrete-event simulation. However, on one hand, specialized simulators, in order to keep an acceptable level of performance, optimize their procedures and data structures for BGP protocol executions; thus, they can not be easily extended to accommodate other routing protocol models. On the other hand, general simulators tailored to investigate the effects of routing protocol dynamics are usually limited to networks of few hundred nodes; thus, large-scale simulations over networks of ten thousands nodes are out of reach. 
\section{Simulator}
\label{simulator}

To measure BGP performances, we rely on the \drmsim ~\cite{drmsim}, a JAVA-based software providing a routing model simulator. \drmsim focuses on the underlying routing layer itself, by exposing a dedicated Application Programming Interface (API) to users. In other words, it is devoted to the construction of routing tables (by means of routing path computation and/or selection) and so to the evaluation of the behavior and performances of various distributed routing models. The main performance metrics supported include the stretch of routing paths produced, the size of routing tables, the number of messages, and the adaptivity to topological modifications.

\subsection{\drmsim architecture}

\begin{figure}[t]
\includegraphics[width=\columnwidth]{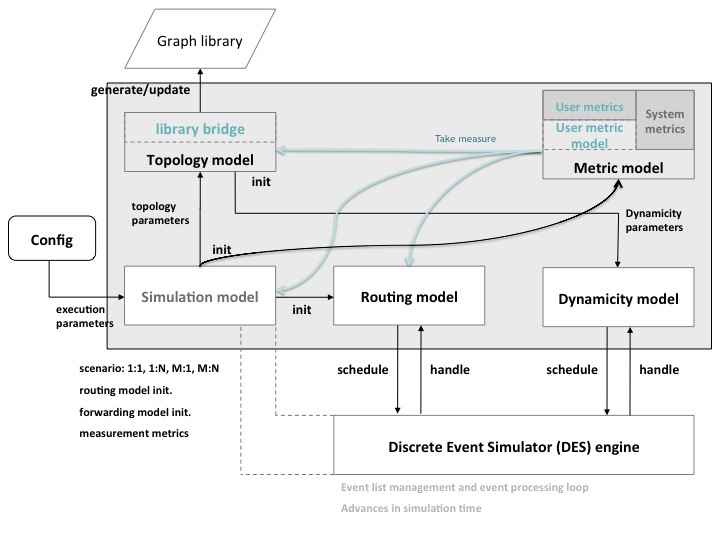}
\caption{\drmsim architecture}
\label{fig_model_arch}
\end{figure}

\drmsim implements the Discrete-Event Simulation (DES) approach. In DES, the operation of a system is represented as a chronological sequence of events (associated with any change in the state of the system). A DES typically implements three data structures: a set of state variables, an event list, and a global clock variable (that denotes the instant on the simulation time axis at which the simulation resides). In \drmsim, an event is a data structure comprising the event's timestamp, the event type, and the event code (a routine which implements what the event consists of). \drmsim comprises a simulation model, a system model, a dynamics model, a metric model and a set of routing models. Figure \ref{fig_model_arch} details the architecture and relationships between these models.

\begin{enumerate}
\item \textit{Simulation model}: initializes the system model, the metric model and the routing model. It also defines the simulation scenario. A scenario could be, for example, the simulation of BGP until convergence upon failure of a set of routers or links initiated at specific times during the simulation.
\item \textit{System model}: controls the network topology. It relies on a graph library to create the network topology, to compute information - like the shortest path matrix - and to perform structural modifications. To avoid dependence on a single graph library and to allow the routing model designer to choose its own graph library, the topology model uses graph library bridges. For each different graph library, a specific bridge must be developed and integrated to \drmsim. If the graph library allows graph partitioning, the system model keeps information about node/link's partition.
\item \textit{Dynamics model}: performs maintenance operations on the network infrastructure as well as router failures. It schedules at a given time - according to the simulation scenario - dynamics events which are router or link failure/repair.
\item \textit{Routing models}: each model comprises the routing procedure(s), the data model and the communication model to be simulated. \drmsim proposes a set of basic routing models (source routing, random schemes, broadcasting, etc). These models allow to verify the correctness of the simulation engine and serve as reference to compare performance with respect to advanced routing protocols. The provided set of models includes the Border Gateway Protocol (BGP) ~\cite{rfc4271}, the Routing Information Protocol (RIP), and compact routing schemes such as NSR ~\cite{Nisse09} and AGMNT ~\cite{Abraham08}.
\item \textit{Metric model}: listens the simulation and topology models. It allows to monitor a selected set of routers/link. This model has been also extended to support measure in case of partitioned network on boundary router/link and partitions. The memory and CPU usage mainly depend on the metrics, on the set of routers/links onto which they are applied, on the measurement interval, and their respective computational complexity. This dependence can leads to extensive use of memory/CPU. To simplify the development of new specific metrics, the metric model is composed of a \textit{system metric model} and a \textit{user metric model}. The former defines a set of default performance related metrics, including the additive routing stretch, the multiplicative routing stretch, the number of routing table entries, and the size of routing tables. The latter provides to the routing model designer, an API to extend the system metric model to perform routing model-specific measures.  
\end{enumerate}

\subsection{Border Gateway Protocol (BGP)}
The Border Gateway Protocol (BGP) is the inter-Autonomous System routing protocol of the Internet. BGP speaking routers exchange network reachability information with other BGP routers on Autonomous Systems (ASs). This information is sufficient for constructing a graph of AS from which routing loops may be pruned, and, at the AS level, some policy decisions may be enforced. Routing information exchanged by means of BGP supports only the destination-based forwarding paradigm, which assumes that intermediate routers forward a packet based solely on the destination address.

\subsubsection{Model description} the network model used in \drmsim considers AS-level topology, meaning that every node represents an Autonomous System (AS). This implies that \drmsim focuses on inter-domain routing, which is today implemented by External BGP (eBGP), i.e., BGP connection between routers in different ASes. Note that eBGP and Internal BGP (iBGP), BGP connection between routers in the same AS, together form the core of the BGP protocol. A router running BGP segments its Routing Information Bases (RIB) into three logical structures for storing routing table entries. First, the Loc-RIB which contains all the selected routes (i.e., a destination prefix, an AS-Path and its associated set of attributes) as locally decided by the router and that are actually used by the forwarding process. Second, the Adj-RIB-In and the Adj-RIB-Out enable the router to provide a neighbor-based filtering for, respectively, incoming and outgoing advertised routes. To simplify processing at each node, a single RIB, the Loc-RIB, is implemented in \drmsim. The Loc-RIB stores routes by taking into consideration the most important attributes (the AS-path and its destination network) while leaving the flexibility for adding new attributes. \drmsim features three implementations of the BGP routing protocol. The first one implements the full BGP state machine. However, because of the large amount of computational resources required to simulate the integral version of BGP, we decided to implement several optimizations.

\subsubsection{Optimizations} in order to reduce the computational resources required for the simulation of BGP, \drmsim implements the following enhancements:
\label{BGP-optimization}
\begin{itemize}
\item Reduction of the number of events by assuming that a BGP session has only two possible states: IDLE or ESTABLISHED. This reduction impacts the establishment time of the BGP sessions. In term of performance, the initial phase will thus complete faster.
\item A router comprises two main data structures: a routing table which contains all the computed/selected routes derived from the routing information received from its peers. This table is usually implemented in software. A router contains also forwarding table which only stores the necessary information to forward packets; it is usually implemented in hardware and, therefore, makes the forwarding process very fast. When simulating routing models, both data structures are coded in software. In order to compare the efficiency of maintaining both data structures or only the routing table, we performed the same simulations using both approaches. We found that maintaining only the routing table data structure and using the "compute on demand" method for the forwarding table entries was the best solution.
\item Code profiling showed that database lookup operations took the largest part of the simulation execution time. Therefore, we investigated many alternatives to overcome this problem. The best solution we found was to assign to each router an identifier from 1 to $n$ (where $n$ is the number of routers in the network) and to index the routing table entries accordingly. The index value for a given routing table entry is the identifier of the destination corresponding to that entry.
\item Adding for every router a bit-vector whose size is the number of routers in our network, and for which the bit at the $i^{th}$ position indicates whether this router has or not a route for the destination with the identifier $i$. Then, efficient logical operations on bit-vector pairs (each composed by the local and the peering router bit- vector) can be performed to determine the useful/useless entries of an update message exchanged between these two routers.
\end{itemize}

\subsubsection{MRAI Impact on BGP Convergence Time}

The dynamics and convergence properties of BGP play an important role in determining network performance as BGP (indirectly) controls the forwarding of inter-domain traffic. Recent studies have shown that the establishment of stable routes after a node failure can take on the order of 3 to 15 minutes \cite{Labovitz00}. 

In a fully connected network, \cite{Labovitz00} demonstrated that the lower bound on BGP convergence time is given by $(N-3)$MRAI, where $N$ is the number of AS in the network, and MRAI is the MinRouteAdvertisementInterval (MRAI) timer \cite{rfc4271}. The MRAI is by default set to 30 seconds on eBGP sessions. This time interval determines the minimum amount of time that must elapse between an advertisement and/or withdrawal of routes to a given destination by a BGP speaker to a peer. Thus, two BGP update messages sent to a given peer by a BGP speaker (that advertises feasible routes and/or withdrawal of infeasible routes to some common set of destinations) are separated by at least one MRAI. This rate limiting mechanism, applied on a per-destination prefix basis, results in suppressing the advertisement of successive updates to a peer for a given prefix until the MRAI timer expires (as it is intended to prevent exchange of transient states between BGP routers). However, the MRAI-based rate limitation results also in routing state coupling between topologically correlated BGP updates for the same destination prefix: the MRAI introduces time synchronization. As a consequence, even if one may think that decreasing the MRAI value would result in decreasing the convergence time, in practice, decreasing the MRAI timer value below a certain threshold leads to adversary effects in terms of number of BGP updates (communication cost) and BGP convergence time ~\cite{Premore01}.

\subsubsection{BGP Metrics}
\label{bgp-metrics}

We are meanly interested in computing the number of BGP update messages. For this purpose, we extended the \drmsim metric model in order to measure the number of update messages, its number of entries and size during all the simulation, but also per router/link.
\subsection{Distributed Parallel Discrete Event Simulation}

When simulating BGP, two main issues appear. First, as already mentioned, storing all routing tables requires a size of $O(k.n^2)$ bits, where $k$ is the size of a routing entry and $n$ is the number of nodes running BGP. Second, storing routing paths received from updates during an MRAI time interval. If a topology comprising 10k nodes can be simulated with an MRAI set to 0s, increasing its value to 30s requires a large amount of additional memory. As a result of this second issue, topology of hundred of nodes can only be simulated. In \cite{SIMROT}, a new data structure is proposed which aggregates shortest paths to reduce their redundancy. If this method can work well for topologies of thousands of nodes, it is still limited for future Internet topologies of hundred of thousand of nodes. 

According to Fujimoto \cite{fuji}, Parallel Discrete Event Simulation (PDES) concerns itself primarily with simulation of asynchronous systems where events are not synchronized by a global clock, i.e., they are truly distributed in nature. It uses hierarchical decomposition of the simulation model to allow an event consisting of several sub-events to be processed concurrently. PDES typically involves utilization of state variables, event list and a global clock variable. The challenging aspect of PDES is to maintain this causality relationship while exploiting inherent parallelism to schedule the jobs faster. On the other hand, extending DRMSim as a PDES requires to extend its engine with precedence graphs so that the engine can retrieve which events can be executed in parallel. Remember that the computational speed-up is expected to range from 0.5 to 2 (as parallel simulation can be slower than sequential one because of synchronization mechanisms). 

Henceforth, moving DRMSim to distributed Parallel Discrete Event Simulation (PDES) seems a promising idea in order to achieve our objective of simulating BGP on large-scale topologies. Indeed, partitioning the topology and affecting each part to a dedicated resource with its own memory/processor communicating through the network, would theoretically divide the required memory by the number of parts. However, several major challenges have to be solved in order to determine the feasibility of such a simulator. 

\subsubsection{Causality errors}
\label{sec:causality_errors}

Let first recall that a DES has \textit{state variables} that describe the simulated system and an \textit{event list} that contains scheduled future events. These events are ordered in the list according to their \textit{timestamp} (occurrence time). An event, whose execution depends on the current state variables, may modify them, add events in the list, and/or remove others that are already scheduled. Hence, if the events are not executed in the right order, the results of the simulation may be wrong. Such errors are called \textit{causality} errors.

We now consider the parallelization of the above paradigm. A PDES is assumed to have many computational resources that share the execution of events. Therefore, causality errors would be the first issue to solve when building a parallel simulator. There are two categories of PDES: \textit{conservative} and \textit{optimistic}. The first one strictly avoids the causality errors while the second one allows them to happen. Of course, the latter needs to detect these errors if they happen and to correct them, which makes its implementation more complicated. 

\begin{figure}[t]
\center
\includegraphics[width=0.7\columnwidth]{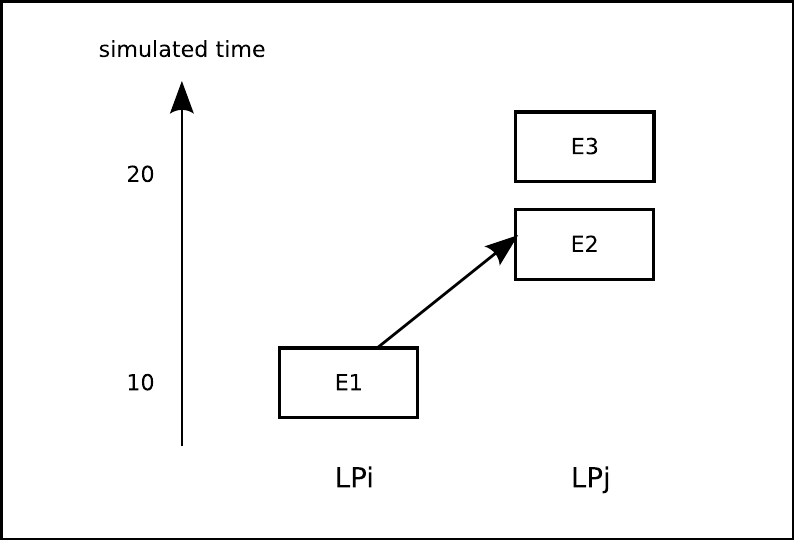}
\caption{Event E1 affects E3 by scheduling a third event E2 which modifies a state variable used by E3. This necessitates sequential execution of all three events}
\label{CausalityError}
\end{figure}

A shared memory containing all state variables, read and written by all the computational resources raises causality errors. These errors are particularly difficult to avoid or handle. Therefore, most PDES strategies prefer to divide the system (the set of routers and peering sessions in our case) between all the computational resources. Each resource only executes events that relate to its part of the system. Formally, the simulator is divided into a set of \textit{logical processes} also called LP. An LP is a sub-simulator which executes, on some computational resource(s), the events related to a subset of the state variables. A logical process LPi might communicate with LPj. If an event E2 is created by LPi that modifies the part of the system that LPj is taking care of, then LPi sends E2 to LPj. This leads to another source of causality errors as shown in Figure \ref{CausalityError}. If LPj didn't consider this possibility, it executes the event E3 directly while provoking a causality error. 

\subsubsection{Deadlocks}

A constraint that is sufficient though not always necessary to guarantee that no causality errors occur is the so-called \textit{local causality constraint}: a discrete event simulation, consisting of logical processes that interact exclusively by exchanging timestamped events, obeys the local causality constraint if and only if each LP processes events in non-decreasing timestamp order. However, when every logical process tries to obey the local causality constraint, and if the implementation is completely decentralized, then it may happen that no LP has enough information to continue, leading thus to a \textit{deadlock} situation.

\subsubsection{Communication cost}
\label{communication_cost}

The logical process methodology requires application programmers to partition the simulator's state variables into a set of disjoint variables. However, the exclusion of shared memory burden some simulations in which an \textbf{interaction} between the partitions exists. In that case, and in the absence of shared state variables, the most natural approach to program such simulations is to "emulate" shared memory by building a logical process for each partition. Each logical processes sending "read" and "write" event messages to access shared information. However this approach often shows poor performances due to message-passing overheads. A more efficient way is to duplicate the shared information in the logical process that needs it; however, in this case, a protocol is required to ensure coherence among the various copies of the shared state variable.

\subsection{Distribution models}

Implementing \drmsim as a distributed parallel discrete event simulator implies that not only routers with their data structures have to be distributed but also that the whole events execution has to be shared among the different computational resources. As stated in Section \ref{communication_cost}, LPs need to communicate between them, thus the simulation feasibility depends mainly on the number of events to be transmitted between each LP and the available bandwidth between them.

Let first describe the main problems we were facing in the non distributed version of \drmsim and how they would be affected by the distribution:
\begin{itemize}
\item Routing tables: let $n$ be the number of routers in the network (providing any-to-any connectivity), and $k$ the the size of a routing entry. The memory needed for storing all the routing tables is in $O(k.n^2)$. This number can reach a large value when $n$ is around tens of thousands. Sharing routing tables among different resources is the main reason behind the distribution of the simulator.
\item Entries updates: this type of event produces the highest number of occurrences. Since for every simulation a new graph is randomly generated, we always start from empty routing tables leading to an initialization phase. It is possible to fill directly the routing tables; however, such operation becomes tedious for some routing protocols that perform, e.g., policy-based filtering or any other rule-based processing not strictly dependent on the topology. 
\end{itemize}

As explained in Section \ref{BGP-optimization}, in the sequential version of \drmsim, the router $R_{source}$ sending an update only add the indexes of these entries from its indexed routing table; on the opposite side, the router $R_{target}$ concerned by this update reads those entries directly from the routing table of $R_{source}$. Such scheme is difficult to maintain in a distributed simulation knowing that $R_{source}$ and $R_{target}$ may not be in the same LP.

The system is distributed among the different logical processes (here, routers and BGP peering sessions). More formally, we model the system by a graph $G(V,E)$, where the set of vertices $V$ is partitioned and every partition is managed by a logical process. This step play an important role. Indeed, as further explained, the performances of the distributed simulations are tightly related to the partitioning algorithm. Edge between two vertices of the same partition is only known by the corresponding LP. For edges with end-points in distinct partitions, we describe in Figure~\ref{edges} two possible solutions. Both studied solutions consider the events update problem:

\begin{figure}[t]
\center
\includegraphics[width=0.8\columnwidth]{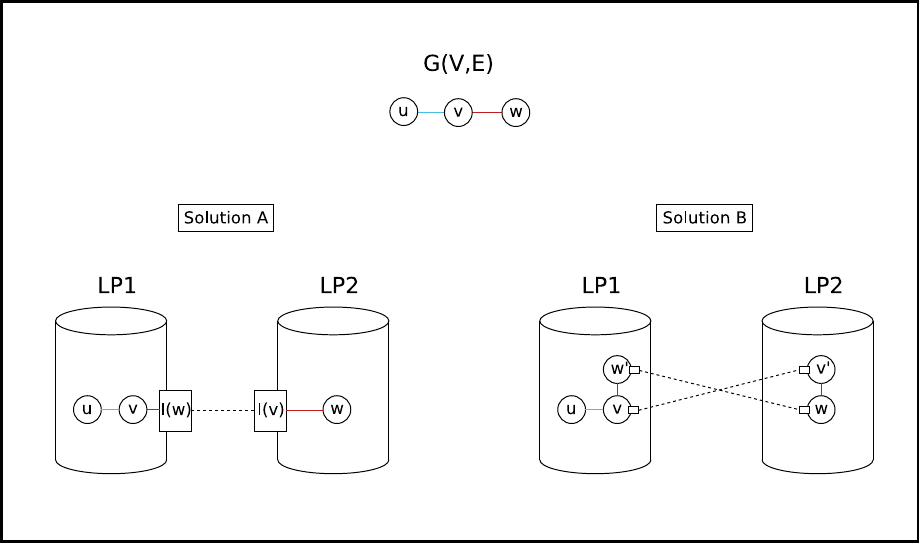}
\caption{Two solutions to allocate the edges which the end-points are on different LPs}
\label{edges}
\end{figure}

\begin{itemize}
\item \textit{Solution A}: the indexes of entries are not sent since $R_{target}$ resides on a different $\text{LP}$ and thus cannot read the entries in the routing table of $R_{source}$. Hence, the update event needs to include the corresponding entries before transmitting them to the LP containing $R_{target}$. In this solution, only the edge needs to be duplicated to handle this \textbf{interaction} event. Note also that we can still use the optimized version of the event when $R_{source}$ and $R_{target}$ are on the same LP. 
\item \textit{Solution B}: we keep the same version of the update event, but we need to duplicate the concerned edge as well as its end-points. This duplication implies that the original vertices (as an example $v$ and $w$ in Figure \ref{edges}) must immediately inform their copies ($v'$ and $w'$) of any change to keep them coherent. 
\end{itemize}

The advantage of Solution B compared to Solution A, is that only modified entries in a routing table are sent, thus reducing the network communication. However, duplication of vertices (routers) may be very harmful if the cuts given by the partitioning algorithm have many edges, reducing in turn the memory gain.
 
To obey the local causality constraint defined in Section \ref{sec:causality_errors}, we use a standard technique based on the following method:
\begin{itemize}
\item For every LP, we consider the number $d$ of \textit{influential} LPs (which may send events to it). Then it needs to keep $d+1$ ordered event lists: one list for the events locally generated, in addition to the list per influential LP to store the received events.
\item A clock is associated to each list. It is equal to either the timestamp of the event at the top of its event list (event with the smallest timestamp) or, in case of an empty list, the timestamp of the last processed event from that list.
\item To select the next event to be processed, an LP chooses the list with the smallest clock. If this list is not empty, the event at the top of the list is processed, otherwise the LP cannot process any event and should wait for new events.
\end{itemize}

\begin{figure}[t]
\center
\includegraphics[width=0.8\columnwidth]{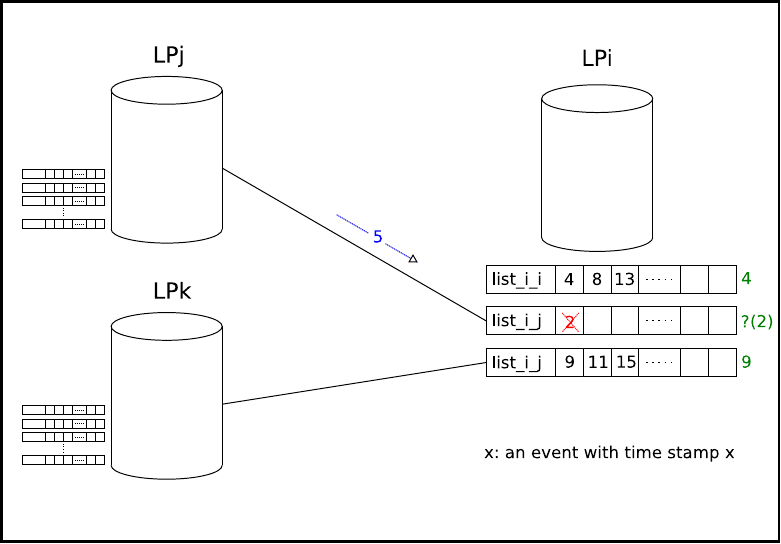}
\caption{An Illustration of the multiple event lists}
\label{LP}
\end{figure}

We give in Figure \ref{LP} an execution example: at the beginning the smallest timestamp in LPi was 2, so its corresponding event was processed. However, since there were no other event in that list, its clock kept the value 2. It was the smallest value among all clocks; so LPi had to pick the next event from this list. But it was empty forcing LPi to wait for new events from LPj. Later, LPj sent an event with timestamp 5 which changed the clock of this list to 5. Now the list of local events have the smallest clock value 4, and it is not empty so LPi can continue. If a cycle of empty lists arises that has sufficiently small clock values, each logical process in that cycle block, and the simulation deadlock. Null events (with empty execution routine) are used to avoid deadlock situations. They do not correspond to any activity in the physical system. A null event with timestamp $T_{null}$ sent from LPi to LPj is just a promise by LPi that it will not send an event to LPj carrying a timestamp smaller than $T_{null}$. One of the possible strategies to avoid deadlocks is to make every LP, whenever it processes an event, send null events to all LPs it may have influence on. But because of the high number of events simulated in \drmsim this solution imply a big amount of network communication, and is thus inappropriate. Another approach consists of sending null events on demand basis rather than after each event. In this case, a longer delay may be required to receive null events as two messages transmissions are required.

\begin{table}[t]
\begin{tabular}{|p{1,2cm}|p{6,8cm}|} 
\hline
Solution A & \textbf{Internal updates}: no communication\\
\medskip
& \textbf{External updates}: 

\begin{center}
$Esize.\sum_{i=1}^k \sum_{v \in V_i} ( |N_{V \setminus V_i}(v)|.|ME(v)| )$
\medskip
\end{center}
\\
\hline
Solution B & 
\textbf{Internal updates}: 
\begin{center}
$Esize.\sum_{i=1}^k \sum_{v \in V_i} [ (\sum_{j=1}^k e(v,V_j)).|ME(v)| ]$
\end{center}
\\
&
\medskip
\textbf{External updates}: the communication here is negligible since we only need to send an update event containing the identifiers of both routers. Even though both LPs will need to process the update event, there is no matter of synchronization and these events can be processed at the same time. \\
\hline
\end{tabular}
\caption{Communications implied by updates}
\label{tab1}
\end{table}

\subsection{Partitioning algorithms and complexity}
\label{partition}

The choice of the partitioning algorithm plays a major role in the PDES performance. A "good" partitioning would evenly distribute processing among LPs (fairness), limit the amount of communication between different LPs and maximize the number of events that can be processed in parallel.

Before introducing the partitioning algorithms, we explain the communication and the memory overhead required by the management of interactions between LPs. We assume that updates constitute the main problem to solve. The other type of events lead to negligible effects. For convenience, we use the following notations:
\begin{itemize} 
\item $G(V,E) \triangleq$ the graph representing the the network;
\item The vertices of the graph are partitioned: $V=\cup_{i=1}^K V_i$;
\item $N_{V'}(v) \triangleq$ the neighbors set of a vertex $v$ on the subgraph induced by $V'$;
\item $ME(v) \triangleq$ the set of modified entries in the router modeled by $v$ ;  
\item $e(v,V')$ =1 if $v$ is not in $V'$ and has a neighbor in $V'$, and 0 otherwise;
\item $Esize \triangleq$ the average size of a routing table entry.
\end{itemize}

Tables \ref{tab1} and \ref{tab2} represent updates induced in terms of communication and memory overhead. Results on implementation of Solution A or Solution B are differentiated. Two types of events may occur: internal updates (blue in Figure \ref{edges}) and external updates (red in Figure \ref{edges}).

\begin{table}[t]
\begin{tabular}{|p{1,2cm}|p{6,8cm}|} 
\hline
Solution A & no memory overhead\\
\hline
Solution B & $Esize.|V|.\sum_{i=1}^k \sum_{v \in V_i} [ (\sum_{i=1}^k e(v,V_j)) ]$ \\
\hline
\end{tabular}
\caption{Memory overhead implied by updates}
\label{tab2}
\end{table}

Among the straightforward techniques to partition graph vertices, one passing through many steps where each of them consists of bipartionning the resulting subsets from the previous step. In this way, partitioning of the graph vertices into $2^q$ subsets would need $q$ steps. We suggest to use mixed integer programs (MIP) to compute these bipartitions. Also a bipartition can be either exact, i.e., both resulting subsets have exactly the same size, or balanced when accepting a difference of $2\varepsilon$ between both sizes. A good bipartition shall minimize the network communication overhead. This minimization problem can be modeled by putting weights on graph components (either on vertices or on edges) and computing a bipartition that minimizes a function of these weights. 

In Solution A, we assign to every edge $uv \in E(G)$ a weight $W_{uv}$ that approximates the number of entries exchanged on this edge. The best bipartition minimizes the sum of weights over all edges having end points in different subsets. Let $(S,\bar S)$ be this bipartition and consider two types of binary variables: 

 \begin{align*}
 \alpha_u &= 
      \begin{cases}
        1,&  \text{if } u\in S\\
    0,&  \text{otherwise}
       \end{cases}
 & \forall u \in V\\
  \beta_{uv} &=
      \begin{cases}
       1,& \text{if } (u,v) \in S\text{x}\bar S \text{~or~} (v,u) \in S\text{x}\bar S\\
       0,&  \text{otherwise}
      \end{cases}
 & \forall uv\in E
  \end{align*}

The following mixed integer program \textbf{MIP1} can be used to compute this optimal bipartition:

\begin{subequations}
\begin{align}
 \min &\ \sum_{uv \in E}  W_{uv} \beta_{uv} &\\
   s.t.\ \  & \alpha_{u} + \alpha_{v} \leq  2 - \beta_{uv}  && \forall uv \in E\\
      & \alpha_{u} + \alpha_{v} \geq  \beta_{uv}  && \forall uv \in E\\
      & \beta_{uv}  \geq |\alpha_{u} - \alpha_{v}|   && \forall uv \in E\\
      & \sum_{u \in V} \alpha_{u} \geq n/2 - \varepsilon \\
      & \sum_{u \in V} \alpha_{u} \leq n/2 + \varepsilon      
\end{align}
\end{subequations}

In Solution B, we assign to every vertex $v \in V(G)$ a weight $W_v$ that approximates $|ME(v)|$. In this case, the best bipartition minimizes the sum of weights over all vertices having at least one neighbor in the foreign subset. Let us add to the binary variables that have been previously introduced a new type of variables.

 \begin{align*}
 \gamma_u &= 
      \begin{cases}
        1,& \text{if } u \text{ has a neighbor in a foreign subset.}\\
    0,& \hspace{.5cm} \text{otherwise}
       \end{cases}
 & \forall u \in V
  \end{align*}
  
The following mixed integer program \textbf{MIP2} can be used to compute the optimal bipartition:

\begin{subequations}
\begin{align}
 \min &\ \sum_{u \in V}  W_{u} \gamma_{u} &\\
   s.t.\ \   & \alpha_{u} + \alpha_{v} \leq  2 - \beta_{uv}  && \forall uv \in E\\
      & \alpha_{u} + \alpha_{v} \geq  \beta_{uv}  && \forall uv \in E\\
      & \beta_{uv}  \geq |\alpha_{u} - \alpha_{v}|   && \forall uv \in E\\
      & \beta_{uv}  \leq \gamma_{u}   && \forall uv \in E\\
      & \beta_{uv}  \leq \gamma_{v}   && \forall uv \in E\\            
      & \sum_{u \in V} \alpha_{u} \geq n/2 - \varepsilon \\
      & \sum_{u \in V} \alpha_{u} \leq n/2 + \varepsilon      
\end{align}
\end{subequations}
\section{Execution}
\label{execution}

Our objective is to determine the communication overhead caused by a distributed parallel implementation of \drmsim 
when simulating BGP. As explained in Section \ref{bgp-metrics}, we are mainly concerned by the number of BGP updates, known to be the most executed event. To compute the number of BGP updates between LPs, we count the number of update messages exchanged at the boundary links of topology partitions. The communication time overhead can be derived by estimating the transmission and the propagation time for a single BGP update event to reach the distributed parallel simulator where it is executed. 

\subsection{Simulation scenarios}

To simulate the BGP model several settings which influence the number of BGP update messages exchanged are considered. For this purpose, we consider 
MRAI = 0s 
in our experiments. It represents an upper bound on the amount of communication between routers for BGP. Indeed, the current value used in the Internet (MRAI=30s) has been set in order to reduce update messages between peers.

The BGP peering session establishment delay together with the update propagation delay between routers play a major role in the amount of transmitted updates. 
For this purpose, three scenarios have been elaborated:
\begin{itemize}
\item \textit{Scenario 1}: considers BGP peering sessions establishment before the start of updates exchanges (BGP emission event). Once sessions are established, received updates are executed in their scheduled order. This scenario simulates a network where all routers have already established BGP sessions with their neighbors. Communication delay of updates between routers is negligible.
\item \textit{Scenario 2}: also considers BGP peering session establishment before the start of updates exchanges (BGP emission event). However, upon reception, updates are executed in a random order simulating highly random communication delay.
\item \textit{Scenario 3}: after one peering session establishment between two routers, the resulting updates are executed. Then, the next scheduled BGP peering session is established. This scenario simulates the arrival of routers one by one in the network, waiting for their convergence before adding a new node.
\end{itemize}

These simulations are executed respectively on topologies of 2.5k, 3k, 3.5k, 4k, 4.5k and 5k nodes. These topologies are generated according to the Generalized Linear Preferential (GLP) model parameterized as specified in \cite{Bu2002}. Each scenario is executed one time for each topologies.

\subsection{Execution environment}


Simulations have been executed on a computer equipped with an Intel Xeon W5580 3.20Ghz with 64GB of RAM (the JVM was limited to 50GB) running 64-bit Fedora 12 on a Linux kernel 2.6.32 together with JRE v1.6.0. The worst simulation times was observed when running the three scenarios on topologies of 5k nodes. Execution of Scenario 1 took 52minutes (min), Scenario 2 48min, and Scenario 16min.

\subsection{Simulation results} 

We first list the number of updates measured by means of the execution environment (see Section \ref{execution}) together with their respective number of entries for simulation of BGP on non-partitioned topologies. These measures provide reference values on the number of updates transiting between routers according to different scenarios. 


\begin{figure}[h]
\includegraphics[width=\columnwidth]{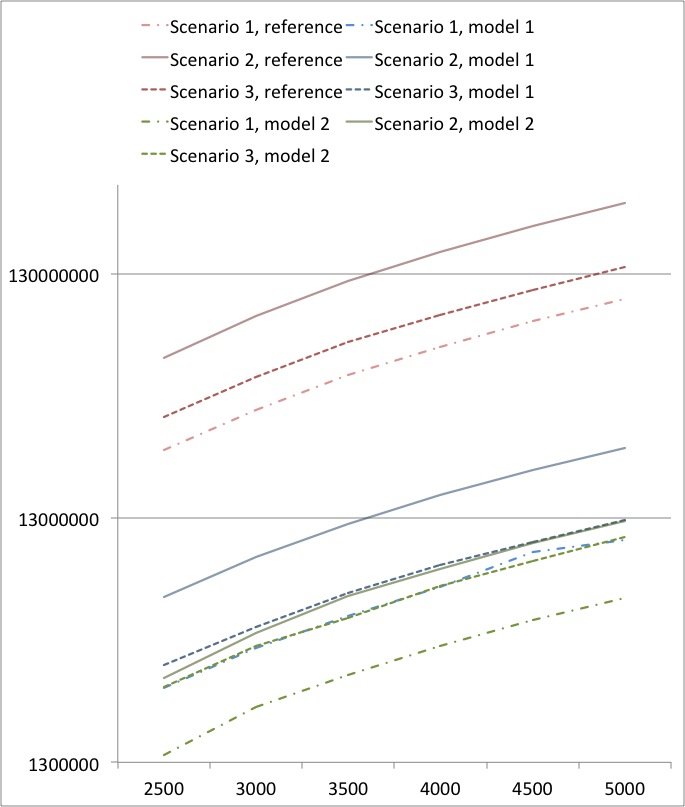}
\caption{Number of update entries for scenario 1, 2 and 3 on non-partitioned and bipartitioned with solution A and B. X axis represents topology size. Y axis the number of update entries}
\label{update}
\end{figure}

\begin{table}[t]
\begin{tabular}{lllllll} 
& 2.5k & 3k & 3.5k & 4k & 4.5k & 5k \\
\hline
\multicolumn{7}{l}{Scenario 1 ($\times 10^6$)}\\
\ No partition & 24,6 & 36,1 & 50,1 & 65,0 & 83,1 & 102,4 \\
\ Sol A on bipartition & 2,6 & 3,8 & 5,1 & 6,7 & 9,4 & 10,5 \\
\ Sol B on bipartition & 1,4 & 2,2 & 2,9 & 3,9 & 5 & 6,1 \\
\hline
\multicolumn{7}{l}{Scenario 2 ($\times 10^6$)}\\
\ No partition & 58,7 & 87,0 & 121,7 & 158,8 & 204,3 & 252,8 \\
\ Sol A on bipartition & 6,1 & 9,0 & 12,3 & 16 & 20,4 & 25,2 \\
\ Sol B on bipartition & 2,8 & 4,4 & 6,2 & 8 & 10,2 & 12,6 \\
\hline
\multicolumn{7}{l}{Scenario 1 ($\times 10^6$)}\\
\ No partition & 33,5 & 49,0 & 68,4 & 88,0 & 111,7 & 138,8 \\
\ Sol A on bipartition & 3,3 & 4,6 & 6,4 & 8,3 & 10,3 & 12,8 \\
\ Sol B on bipartition & 2,6 & 3,9 & 5 & 6,9 & 8,6 & 10,9 \\
\hline
\end{tabular}
\caption{Number of update entries}
\label{non-partitioned-table}
\end{table}

%
%

\subsubsection{BGP reference results on non partitioned topologies} 
\label{non-partitioned}

From Figure \ref{update} on non partitioned topologies, the number of BGP update entries increases drastically (up to $10^8$ update entries). If the communication overhead between LPs behaves similarly, then the time to perform simulations becomes too high. 
This observation corroborates the fact that setting MRAI value to 0s leads to detrimental effects in terms of routing convergence time.
We also observe that the increase of BGP update entries over the topology size is linear in their root square, allowing us to extrapolate this number for the topology size of interest (from 40k to more than 100k nodes). The general behavior of these observations has to be further confirmed when considering only the boundary links on partitioned topologies.

Scenario 1 is the most effective in terms of number of updates entries
 compared to the other scenarios. 
 Scenario 2 is the less effective in terms of number of updates 
  compared to the other scenarios. These results can be explained by the dependence on the spatial concentration of the initiation BGP update events in a large-scale topology. Scenario 1 represents the situation where perturbation is characterized by a single event at a time that represents the perturbation of a global stable state by a single state update at a time processed as it arrives. 
  Scenario 2, equivalent to a delayed synchronization between all router pairs, shows the detrimental effect of increasing processing delays of updates where individual events can randomly induce a routing table change across the topology at random time scales and events interfere with each other. In Scenario 3 (equivalent to the addition or a router in the topology) perturbations are characterized by a set of events whose initiation is spatially concentrated (instead of a set of events spatially distributed over the entire topology like Scenario 2). Scenario 2 and 3 lead to results of the same order of magnitude; 

\subsubsection{BGP results on bipartitioned topologies with Solution A}
\label{partitioned}

Let now consider bipartitions of our topologies. We compute optimal bipartitions according to the methodology presented in Section \ref{partition} with Solution A. For this purpose, we provide as edge weight the reference number of update entries measured previously. As optimal bipartitions, we know that the measured communication on boundary edges cannot be lower with any other partitioning algorithms, giving us a lower bound on the overhead. We re-execute the previous experiments and compute only the number of updates (together with their entries) transmitted between the partitions. In Figure \ref{update}, we plot the measured number of update entries for each scenario. 

As expected from Section \ref{non-partitioned}, for the three scenarios and a topology of 5k routers, around 10\% of the total number of updates entries transit between partitions. Table \ref{non-partitioned-table} shows the measured number of entries.
 
\subsubsection{BGP result on bipartitioned topologies with Solution B}

In Section \ref{partition}, we present a second model of distribution, named Solution B, that reduces the communication needed between partitions compared to Solution A. If a router in partition 1 has boundary neighbors not in the same partition than partition 1, only one message between partition 1 and boundary neighbor's partition is necessary. We recompute the optimal bipartitions according to this second model. Edge weight was again took from the reference number of update entries measured previously. As optimal bipartitions, we know that the measured communication on boundary edges cannot be lower with any other partitioning algorithms. Here, we compute only the number of updates (together with their entries) transmitted between the partitions. Figure \ref{update} shows the measured number of update entries for each scenario with Solution B.

We observe for the three scenarios improvement on the number of transmitted entries. Using Solution B, only 5,6\% of the reference update entries have to be transmitted with scenario 1, 5\% for Scenario 2 and 8\% for Scenario 3. Table \ref{non-partitioned-table} shows the measured number of entries.
 
\subsubsection{Communication overhead}

From Section \ref{update}, we know how many update entries have been transmitted between two partitions, by computing the average size of an update entry we are able to compute the amount of communication between LPs. We measured the total size of update entries transmitted for the different scenarios and topologies. Except for Scenario 1, where the average size of entries is 1, the other scenarios produce an average entries size between 2,8 and 4,5. Note that in the case of BGP, this quantity provides an estimation of the average size of the shortest paths among all topologies.

To compute the overhead time of a distributed parallel discrete-event simulator, we also need to know how long it takes to transmit an update entry between two LPs. Let consider two LPs communicating through our local network at INRIA Labs. Each LP is executed on different computers connected through Ethernet interfaces of 1Gbps. We relied on TCP/IP and Java sockets to transmit information between the LPs. We measured the time (including the waiting time of the ACK of TCP packet) to transmit an update entry between the two computers. An update entry is considered as a TCP packet containing one integer. One packet sent from one computer took 2,6ms to reach its destination. By sending 1000 packets one after the other, we obtained an average of 0,26ms/packet to reach its destination. We saw previously that the size of the update entries can vary between 1 and 4,5. However, as soon as the average size of the update entries does not exceed the default size of a TCP packet, no impact on the transmission delay is noticed. Thus, we only consider the 0,26ms/packet transmission time. 

With Solution A applied to Scenario 1 on a topology of 5k routers, there are $10.10^6$ entries to transmit, so we can expect 2745s (about 45min) to transmit all update entries between the LPs. As the measured number of update entries and their size are linear in their square root from one topology size to another, we can expect around 5490s (about 91min) of overhead for a topology of 10k routers and 54900s (about 15hours) for a topology of 100k routers. Considering the two other scenarios and a topology of 5k routers, Scenario 2 has $25.10^6$ entries and cumulates an overhead of 6561s (110min), and Scenario 3 with $12.10^6$ entries stays in the same order of time than Scenario 1.

If we take the highest value, we would cumulate an overhead of 6561s to transmit all update entries between the LPs. The measured update entries are also linear in their square root from one topology size to another for both scenarios, so we can expect around 13122s of overhead for a topology of 10k routers and 131220s (about 36hours) for a topology of 100k routers. These results show that scenarios where perturbations are spatially distributed and the number of perturbed states limited compared to the total number of routing states can be accommodated with reasonable time overhead. On the contrary, scenarios constructed on spatial randomization of small perturbations (Scenario 2) or large number of small perturbations (Scenario 3) significantly increase this overhead. Assuming that nodes would be interconnected by 10Gbps links, one would still reach about 3 hours of cumulated time overhead for a topology of 100k routers. With Solution B applied to Scenarios~1 and~2, the number of update entries is reduced by a factor of 2 compared to Solution~A. Scenario 3 allows less improvement as only 2\% of the entries are saved.

\section{Conclusion}
\label{conclusion}

A first step in the investigation of new routing models for the Internet has been reached by enabling simulation of the Border Gateway Protocol (BGP) on topologies of order of $O(10k)$ nodes as well as by comparing it new experimental routing models. However, the expansion of Internet and its dynamics require not only to go a step further in terms of topology size but also to be able to handle the evolutions of the Internet routing system for the next decade. Moving our routing model simulator \drmsim to Distributed Parallel Discrete Event seems to provide a promising technique in order to make abstraction of the size of the topologies, but at the cost of an induced communication overhead between logical processes. 

The main objective of this paper is to quantify the expected additional time needed to simulate BGP on such a distributed architecture for topology sizes of interest, i.e., 10 to 100k. For this purpose, we have identified BGP updates as the main cause of communication between nodes and computed the number of updates as well as the number of entries composing an update for different BGP execution scenarios. Then, we have computed the best bipartition - based on the previous measured number of updates - of each topology to derive the amount of communication between the partitions and so on the overhead of the distributed simulation. It appears that for Scenario 1 (corresponding to the case where all routers their BGP peering sessions established before the start of updates exchanges) additional time ranging from 91min (topology of 10k nodes) to 15hours (topology of 100k nodes) is required to perform a distributed simulation. As shown by the Scenario 2 (Scenario 1 with random updates) and Scenario 3 (corresponding to the addition of routers one-by-one in the topology), this additional time clearly depends of the execution scenario and can reach hours up to 36hours for topologies of 100k nodes. Decreasing the number of updates can be achieved by increasing the MRAI time to a value higher than 0 (as the MRAI aims at reducing the BGP update rate). Indeed as observed in \cite{Premore01}, there is no rate limiting when MRAI=0 and a large number of updates are needed for BGP convergence. As the MRAI increases, the number of BGP updates decreases significantly (in the order of factor 10 depending on the reference value), until the convergence time reaches an average minimum corresponding to an optimal MRAI value. After this value, the convergence time only increases while the total number of updates doesn't vary significantly anymore. In the context of this paper, the critical issue becomes thus to determine the value of the MRAI for which the decrease in update rate wouldn't increase the convergence time while limiting the amount of additional memory required by the simulator. In order to further reduce the overhead, we could consider more advanced techniques such as the recently introduced Path Exploration Damping (PED) \cite{huston10}; however using this technique would modify the usual BGP router's model. Instead, we have proposed to synchronize boundary nodes between partitions. By applying this method, we were able to save half of the overtime when considering Scenario 1 and 2. Obviously, such a distributed parallel simulator seems thus feasible leading the next step of our work to be the extension of our routing model simulator \drmsim according to the proposed distribution models.

\section*{Acknowledgments}
This work has been partially supported FP7 STREP EULER.

\end{document}